# Vibrations of Dimers of Mechanically Coupled Nanostructures : Analytical and Numerical Modeling

Jean Lermé, Jérémie Margueritat, and Aurélien Crut[*]

*Université de Lyon, CNRS, Université Claude Bernard Lyon 1, Institut Lumière Matière, F-69622 Villeurbanne, France*

**ABSTRACT**

The coupling effects affecting the vibrations of two close nanostructures (e.g., two metal nanoplates or nanospheres separated by a thin dielectric layer) may considerably alter their vibrational eigenfrequencies, as demonstrated by several recent experimental studies. In this work, we present theoretical investigations of these coupling processes based on a continuum mechanics approach, considering various systems composed by two identical nanostructures mechanically coupled by a spacer made of a different material and computing their eigenfrequencies as a function of the spacer thickness. We first discuss the vibrations of stacked slabs, a one-dimensional problem which can be treated analytically. The more complex configurations of dimers of rods or spheres coupled by a finite cylindrical spacer are then treated numerically. In all cases, the frequency shifts occuring for thin spacers can be simply interpreted as a modification of the boundary conditions of the problem as compared to the single nanostructure case, while those predicted near specific spacer thicknesses are ascribed to an avoided crossing effect, happening when the individual building blocks of the dimers (nanostructures and spacer) present common eigenfrequencies.

**INTRODUCTION**

Nano-objects present discrete acoustic vibrational modes, whose intrinsic features (frequency and damping rate) are set by the morphological and elastic properties of the nano-objects and their environment. Measuring the vibrational response of nano-objects thus constitutes a powerful tool to characterize their size, shape or composition. From an experimental point of view, the vibrations of nano-objects can be detected using optical spectroscopy techniques working in the time (pump-probe time-resolved spectroscopy) or frequency (Raman/Brillouin scattering spectroscopies) domains.[1–6]

These techniques typically allow the detection of one or a few nano-object modes with frequencies in the GHz-THz range, of the order of the ratio between sound velocity and nano-object size. The experimental and theoretical investigations performed in the last twenty years have largely clarified the dependence of the vibrational frequencies of isolated nano-objects on their size, shape and elastic properties,[7–12] and shown that they are usually reliably predicted by continuum mechanics approaches for nano-objects with sizes down to ≈1 nm, without having to consider a size-induced modification of their elastic constants.[4,5,13–16] In this context, the interest of nanoacoustics researchers is now shifting to less understood aspects of the vibrational response of nano-objects, including in particular damping and coupling phenomena, the latter constituting the scope of this paper.

Vibrational coupling between close nano-objects takes place when they are mechanically connected, i.e. when the vibrations generated in a nano-object may reach close ones by propagation through the medium separating them. This situation may occur in a variety of systems, differing by the number of nano-objects (e.g., nanoparticle dimer, oligomer or supracrystal) and the nature and strength of their mechanical connections (e.g., *via* surfactant molecules, a polymer matrix or a supporting solid substrate).[17–27] For metallic nano-objects, such proximity also simultaneously generates plasmonic interactions, which strongly affect their optical response (redshifting for instance the surface plasmon resonances of plasmonic homodimers as compared to those of isolated nanoparticles).[28–32] This effect offers the possibility to selectively probe dimers in optical spectroscopy experiments involving ensembles of nanoparticles, by tuning the light wavelength with a plasmonic resonance generated by plasmonic interactions.[24,25,33] The high sensitivity of the optical response of dimers to interparticle distance also makes their inelastic light scattering spectra much richer than those of isolated nano-objects.[33]

The acoustic interactions between close supported nano-objects only coupled by their underlying substrate are typically weak, and initial experiments performed on nanodimers formed by two close non-touching nano-objects (pairs of nanoprisms[34] and nanocubes[35] lithographed on a substrate) showed no clear signature of coupling effects. Nevertheless, coupling-induced frequency shifts were later observed in more complex nanoparticle oligomers (decamer of gold nanodisks, where a central nanodisk was surrounded by an outer ring of nine smaller nanodisks).[26]

In contrast, multiple coupling effects were observed in time-resolved and inelastic light scattering experiments involving dimers of nano-objects either in direct contact or separated by a nanometric organic layer.[19,21,24,25,27,36] In particular, measurements on samples containing nanosphere dimers[21,24,25] allowed the detection of a low-frequency vibrational mode, associated with periodical dimer stretching by opposite quasi-translational motions of the two nanospheres, whose frequency was shown by numerical simulations to be much higher (almost double) than that associated with the quasi-translational motion of an isolated matrix-embedded nanosphere (labeled ($l$ = 1, $n$ = 0) in the framework of Lamb theory).[25] The reproduction of inelastic scattering experiments on single dimers enabled finer information to be extracted, allowing to distinguish the contributions of two distinct vibrational modes in the low-frequency range of inelastic scattering spectra, which were ascribed to coupled motions of the nanoparticles parallel or orthogonal to the dimer axis.[36] These experiments also demonstrated a large (≈25%) coupling-induced frequency increase for the quadrupolar mode ($l$ = 2, $n$ = 0), which dominates the inelastic scattering spectra of isolated nanospheres.[2,33,37] Observations of similar vibrational coupling effects were made in recent time-resolved investigations on single dimers of gold nanoplates separated by a thin PVP layer. The stacking of two nanoplates with close ≈50 GHz breathing frequencies[19] was seen to produce two detectable eigenmodes, with frequencies respectively similar to the initial ones and 15% higher, in excellent agreement with the predictions of a simple coupled spring model. A low-frequency mode ascribed to a relative motion of stacked plates was also detected.

Plasmonic coupling has been largely investigated from a theoretical point of view, a fruitful analogy between the hybridization of plasmonic modes in a dimer of nanoparticles and that of atomic orbitals in a molecule having in particular been presented.[28,38] In contrast, vibrational coupling has been much less explored. Up to now, only a few modeling attempts have been reported, based either on empirical coupled spring models or on the use of fully numerical calculations.[19,22,25–27,36] The goal of this theoretical paper is to clarify the vibrational interactions occurring in a dimer of nano-objects, by combining analytical tools, which enable an equation-based discussion of coupling phenomena for simple systems, and numerical ones which allow addressing more complex geometries, using in both cases the framework of continuum mechanics. We considered here systems involving a pair of identical nanostructures separated by a finite spacer (the names of « *nanostructures* » and « *spacer* » being used to designate them throughout the paper), assuming a purely elastic response for both materials. The absence of damping for the vibrational modes of such systems facilitates physical interpretations (e.g., regarding mode crossing) as compared to the systems including an infinite environment, in which case vibrational modes are damped because of the emission of acoustic waves in the environment.[25,36] We first consider the vibrational response of a dimer of two identical slabs with infinite lateral extension separated by a thin spacer, a problem which can be treated analytically due to its one-dimensional character, allowing equation-based physical interpretations to be found in both the small and large separation regimes. We then numerically investigate the vibrations of another system related to the previous one by its quasi-one-dimensional character (two nanorods with length much larger than diameter connected by an intermediate cylinder of same radius), and show that they are governed by the same phenomena. Finally, we numerically address the more complex case of a dumbbell made of two nanospheres connected by a cylindrical spacer, and find that its vibrations can still be understood using the same considerations as previously. The case of a dimer of gold nanostructures connected by a polyvinylpyrrolidone (PVP) spacer was taken as a reference for all these calculations because of its experimental relevance. Indeed, this situation was encountered in two recent series of experiments which addressed vibrational coupling between pairs of gold nanospheres[25,36] and nanoplates[19,27] separated by PVP.

**METHODS**

**Continuum mechanics approach.** Only elastically isotropic materials were considered in this work. Their mechanical properties depend on only three parameters : their density $\rho$ and a pair of elastic constants, which can be for instance the Lamé constants $\lambda$ and $\mu$, the Young modulus $Y=\mu(3\lambda+2\mu)/(\lambda+\mu)$ and Poisson ratio $\nu=\lambda/(2(\lambda+\mu))$, or the longitudinal and transverse sound velocities $c_L=((\lambda+2\mu)/\rho)^{1/2}$ and $c_T=(\mu/\rho)^{1/2}$ (prime notations being used for the spacer in the following). For such materials, the displacement field **u** obeys Navier equation, which writes

$$\rho \frac{\partial^2 \mathbf{u}(\mathbf{r},t)}{\partial t^2} = (\lambda + 2\mu)\nabla(\nabla \cdot \mathbf{u}) - \mu \nabla \times (\nabla \times \mathbf{u}) \qquad (1)$$

Perfect mechanical contact (i.e., continuity of displacement and stress) was assumed here at all interfaces between different media. Vibrational modes correspond to harmonic solutions of eq 1 $\mathbf{u}(\mathbf{r},t) = \mathbf{u}(\mathbf{r})e^{-i\omega t}$. They depend on the boundary conditions (BCs) specified at the border of the system and are characterized by their angular frequency $\omega=2\pi f$ (real for finite systems in the absence of internal dissipation processes) and their associated displacement field $\mathbf{u}(\mathbf{r})$.

**Numerical models.** The numerical simulations presented in this work were performed using the Structural Mechanics module of the COMSOL finite-element modeling commercial software. The axisymmetric vibrational eigenmodes of isolated and coupled nanostructures were computed using the boundary conditions indicated in the text, using an Eigenfrequency calculation type.[39] Geometrical symmetries were exploited to perform calculations in 2D rather than in 3D, as discussed below. About 10000 mesh elements were used for each calculation.

**Parameters used in the simulations.** $Y$=79 GPa, $v$=0.44 and $\rho$=19300 kg.m$^{-3}$ were used for the Young modulus, Poisson ratio and density of gold nanostructures (corresponding to $c_L$=3642 m.s$^{-1}$ longitudinal sound velocity and $Z=\rho c_L$=70.3 10$^6$ kg.m$^{-2}$.s$^{-1}$ acoustic impedance), while $Y'$=4 GPa, $v'$=0.39 and $\rho'$=1200 kg.m$^{-3}$ values (corresponding to $c_L'$= 2580 m.s$^{-1}$ and $Z'$=3.1 10$^6$ kg.m$^{-2}$.s$^{-1}$) were used for PVP spacers.

**RESULTS AND DISCUSSION**

a) **Vibrations of coupled slabs**

We first consider the longitudinal vibrations of two identical slabs of thickness $h$ and infinite lateral extension (nanostructures), mechanically coupled by an intermediate slab of thickness $d$ (spacer), as shown in Fig. 1a. We note $z$ the longitudinal coordinate, **u$_z$** the corresponding unit vector and choose the origin of the axis O in the symmetry plane of the system, so that the spacer extends from $z$=-$d$/2 to $z$=$d$/2, and the gold ones from $z$=±$d$/2 to $z$=±($d$/2+$h$) (Fig. 1a). Stress-free BCs are considered at the external surfaces of this system, which corresponds to analyzing its vibrations in vacuum. This one-dimensional problem can be treated analytically (as fully detailed in the Supporting Information). For the purely longitudinal motions considered here (**u**=$u(z)$ **u$_z$**), Navier equation (eq 1) reduces in each slab to

$$\frac{\partial^2 u(z,t)}{\partial t^2} = c_L^2 \frac{\partial^2 u(z,t)}{\partial z^2} \qquad (2)$$

Therefore, displacement and stress fields are in each slab of the form $u(z) = A\cos(\frac{\omega z}{c_L}) + B\sin(\frac{\omega z}{c_L})$ and $t(z) = \rho c_L^2 \frac{\partial u(z)}{\partial z}$, respectively, the $A$ and $B$ coefficients in adjacent slabs being related by continuity relations, due to the assumption of a perfect mechanical contact at internal interfaces (see the Supporting Information for details). Due to the symmetry about the $z$=0 plane of the considered problem, the displacement field associated with vibrational modes is either symmetric ($u$(-$z$)=-$u$($z$), with thus $A$=0 in the spacer) or antisymmetric ($u$(-$z$)=$u$($z$), with thus $B$=0 in the spacer) relatively to the $z$=0 plane.

The frequencies of eigenmodes are fixed by the stress-free BCs at the external end of the nanostructures ($t(\pm(d/2+h))=0$). Introducing dimensionless frequencies defined as $\xi = \dfrac{\omega h}{c_L}$ (vibrational periods being proportional to the size of the system in the context of continuum mechanics) and $p = \dfrac{\rho' c_L'}{\rho c_L} = \dfrac{Z'}{Z}$ the ratio of the acoustic impedances of the spacer and of the nanostructures, frequencies are given for symmetric modes by

$$p \cos\left(\frac{c_L}{c_L'}\frac{d}{2h}\xi\right)\cos(\xi) - \sin\left(\frac{c_L}{c_L'}\frac{d}{2h}\xi\right)\sin(\xi) = 0 \qquad (3)$$

and for antisymmetric ones by

$$p \sin\left(\frac{c_L}{c_L'}\frac{d}{2h}\xi\right)\cos(\xi) + \cos\left(\frac{c_L}{c_L'}\frac{d}{2h}\xi\right)\sin(\xi) = 0 \qquad (4)$$

Fig. 1b presents the lowest reduced vibrational frequencies $\xi/\pi=2fh/c_L$ computed for the considered Au-PVP three-slabs system (Fig. 1a) as a function of the dimensionless separation parameter $d/h$ appearing in eqs 3 and 4 (red and black lines corresponding to symmetric and antisymmetric modes, respectively). In order to better understand the evolution with $d/h$ of the vibrational frequencies of the full Au-PVP system and the way they originate from inter-slabs mechanical coupling effects, it is useful to compare them with those associated with the individual components of the dimers. In the case of gold nanostructures separated by a PVP spacer, the acoustic impedance ratio $p$ value at the nanostructure/spacer interface is small ($p\approx0.044$), so that the relevant frequencies of isolated nanostructures and spacers to be considered are those obtained by taking the $p\rightarrow0$ limit in eqs 3 and 4. They correspond to decoupled vibrations of the nanostructure and spacer slabs with stress-free and displacement-free BCs on the slab surfaces, respectively ; details on the vibrations of isolated slabs are provided in the Supporting Information. The frequencies of an isolated gold slab with stress-free BCs are shown as horizontal dashed blue lines in Fig. 1b, and correspond to $\sin(\xi)=0$, i.e. integer values $n$ of the used reduced frequency $\xi/\pi$, indicated at the right of Fig. 1b. Those computed for an isolated PVP slab of variable thickness $d$ with displacement-free BCs are shown as magenta and gray dashed lines, for symmetric and antisymmetric modes, respectively (corresponding to the conditions $\sin(\omega d/(2c_L'))=0$ and $\cos(\omega d/(2c_L'))=0$, respectively).

Figure 1b shows that the eigenfrequencies of the three-slabs system form continuous, non-crossing branches. All branches present a similar step-like evolution, consisting of an alternation of parts where eigenfrequencies are decaying with $d/h$ and parts where they remain almost constant. In most cases, the eigenfrequencies of the full Au-PVP system are close to those of isolated Au slabs (in the horizontal step parts of the branches) or PVP slabs (in the decaying parts). In particular, the frequencies of isolated gold slabs are generally (i.e., for most $d/h$ values) approximately retrieved, along the horizontal parts of the branches, by one symmetric mode and one antisymmetric mode of the three-slabs system, the displacement field in the gold component being very similar for the coupled and uncoupled modes of close frequencies, as demonstrated for instance by the hybridization analysis performed in part D of the Supporting Information, in which a comparison with plasmon hybridization is also presented. Therefore, for the considered Au-PVP three-slabs system, the main effect of vibrational coupling is usually a splitting of each nanostructure mode into two modes with close frequencies, corresponding to symmetric and antisymmetric combinations of the nanostructure modes (according to the symmetry character defined previously). Such splitting is a very general phenomenon occurring when identical oscillators are coupled, already observable in the simplest coupled pendulum/spring models.[40]

The good overall match between the eigenfrequencies of the three-slabs Au-PVP systems and those of their individual components can be ascribed to the large mismatch between the Au and PVP acoustic impedances, which leads to moderate coupling effects, reducing in most cases to the splitting of gold slab modes discussed above, with frequencies close to those obtained in the $p \rightarrow 0$ limit. A more general discussion of the dependence on $p$ of the eigenfrequencies of three-slabs systems can be found in part E of the Supporting Information. In particular, the larger deviations occurring for a reduced acoustic mismatch between the nanostructure and spacer materials are illustrated by Fig. S8 of the Supporting Information, in which the spacer density $\rho'$ was numerically modified to yield $p$ values of 0.18 and 0.55. However, even in the Au-PVP case, marked deviations between the eigenfrequencies of dimerized and isolated gold slabs occur in two distinct cases : near points where the eigenfrequencies of the isolated Au and PVP slabs cross each other, and in the quasi-contact regime ($d/h \rightarrow 0$).

An avoided crossing behavior is observed in Fig. 1b in the eigenfrequency branch pattern of the Au-PVP three-slabs system at all ($d/h$, $\xi$) points where one of the spacer eigenfrequencies, which decay with $d/h$, equals that of a nanostructure mode (independent of $d/h$). Mathematically, the impossibility for an eigenmode of the Au-PVP system to have the same frequency as a spacer mode of same symmetry/antisymmetry character at the $d/h$ value when it coincides with a nanostructure eigenfrequency is a clear consequence of eqs 3 and 4. For instance, eq 3 cannot be satisfied if $\sin(\xi)=0$ and $\sin(\omega d/(2c_L'))=0$ simultaneously, as the left-hand term of the equation (neglected in the $p \rightarrow 0$ limit) would equal $p$ in this case. From a more physical perspective, this avoided crossing behavior results from the different BCs (stress-free vs displacement free) associated to the nanostructure and spacer modes shown in Fig. 1b, which cannot be juxtaposed to form a coupled mode of same frequency. Note that a similar argument was previously used to understand the frequencies of the radial modes of core-shell spherical nanoparticles detected in the context of time-resolved experiments.[41,42]

In the quasi-contact regime ($d/h \to 0$), the crossing effects discussed in the previous paragraph cease to affect low-frequency modes of the Au-PVP system as the frequency of spacer modes tends to infinity. However, the symmetric and antisymmetric modes to which each gold slab mode gives rise present increasingly different eigenfrequencies. Indeed, $\xi/\pi$ shows an increase of 1/2 as $d/h$ is decreased (on a range decreasing with frequency) for symmetric branches, while for antisymmetric ones it remains almost constant. In the situation considered here, vibrational coupling effects are thus larger for symmetric combinations of low-frequency modes, leading in particular to a 50% frequency increase for the $n = 1$ (breathing) mode. The $d/h \to 0$ limit can be understood by making $d/h=0$ in eqs 3 and 4, which then reduce to $\cos(\xi)=0$ and $\sin(\xi)=0$, respectively, or equivalently $\sin(2\xi)=0$. The solutions $\xi=(n+1/2)\pi$ and $\xi=n\pi$ correspond to the symmetric and antisymmetric modes of a gold slab of thickness $2h$ with stress-free BCs. The solutions for the symmetric coupled modes, namely $\xi=(n+1/2)\pi$, also correspond to the vibrational modes of an individual Au slab of thickness $h$ with stress-free BCs at one end and displacement-free ones at the other end (see the Supporting Information). Therefore, the increased symmetric mode frequencies of two identical slabs induced by their increased coupling through an intermediate slab with decreasing thickness can be qualitatively interpreted as the signature of a progressive change of BCs at one end of the slab, which has a deep impact on its displacement field (as shown by the hybridization analysis presented in Fig. S7 of the Supporting Information). The frequencies obtained in the contact case ($d/h=0$) are thus independent of $p$ for both symmetric and antisymmetric modes. However, their values for small but non-vanishing $d/h$ strongly depend on the spacer parameters (thickness, density and sound velocity). This dependence is illustrated in Fig. S9 of the Supporting Information in the experimentally relevant case of the fundamental ($n = 1$) breathing mode of gold slabs, where both PVP spacers and heavier ones (leading to reduced acoustic impedance mismatch) were considered. It can also be highlighted by a first order expansion of eqs 3 and 4 when $d/h \to 0$, which leads to the approximate reduced eigenfrequencies

$$\frac{\xi}{\pi} = \frac{2fh}{c_L} = (n+\frac{1}{2})\,(1-\frac{c_L}{2c_L'}\frac{d}{h}\frac{1}{p}) \qquad (5)$$

for symmetric modes and

$$\frac{\xi}{\pi} = \frac{2fh}{c_L} = n\left(1 - \frac{c_L}{2c_L'}\frac{d}{h}p\right) \tag{6}$$

for antisymmetric ones.

The simple slab model presented here can be applied to the analysis of recent vibrational measurements performed on gold nanoplates separated by a thin PVP layer.[19,27] Indeed, the lateral size of the nanoplates used in these studies (10-20 μm) was much larger than their thickness (20-50 nm), making the use of a 1D model for analyzing their vibrations meaningful. In these works, optical pump-probe measurements were performed on single dimers of two partly overlapping stacked nanoplates. Close ≈50 GHz $f_1$ and $f_2$ frequencies were detected for pump and probe beams focused in non-overlapping regions, and ascribed to the fundamental ($n$ = 1) frequencies of the two individual nanoplates ($f_1$ and $f_2$ being slightly different in most cases, which presumably originates from the thickness dispersion of the nanoplates). In overlap regions, two different vibrational frequencies $f_-$ (between $f_1$ and $f_2$) and $f_+$ (15% higher than $f_-$ for measurements in which $f_1$ and $f_2$ were almost identical) were simultaneously detected. These observations were successfully analyzed through an empirical coupled spring model, in which the coupling strength between the nanoplates was left as a free parameter. They can be further rationalized by our slab model, which predicts in the near-contact regime the splitting of each nanostructure mode into an antisymmetric dimer mode with almost unchanged frequency (as $f_-$) and a symmetric one with increased frequency (as $f_+$). Moreover, a symmetric Au-PVP mode with a frequency 15% larger than that of the $n$ = 1 Au slab mode is obtained in our model when $d/h$≈0.03 (Fig. S9), which corresponds to a plausible $d$≈1 nm PVP thickness for a $h$=30 nm thick gold nanoplate. Indeed, such ≈1 nm PVP surfactant layer thickness is typical in the context of metal nanoparticle synthesis.[43] Note however that the deduced $d$ value is just an estimation, because vibrational frequencies may be affected by the deposition of the nanoplates on a supporting substrate, which is not included in our simple model. The additional low-frequency mode (with a frequency of about 20 GHz, i.e. 2.5 times lower than $f_1$ and $f_2$) detected in the overlap region can be ascribed to the lowest-frequency symmetric mode of the Au-PVP system mode, to which the translation mode ($n$ = 0) of the gold slab gives rise, which has a reduced frequency of $\xi/\pi$≈0.35 (i.e., about 0.35 $f_1$, as experimentally observed) for the $d/h$ value of 0.03 deduced above in the framework of our model (Fig. 1b). Therefore, our simple model appears to be able to provide a quantitative description of the mechanical coupling effects arising in these stacked plates systems.

An analytical analysis of vibrational coupling such as that presented in this part is however possible only for highly symmetric systems, being straightforward only when the calculation of coupled vibrational modes reduces to a 1D problem. Numerical methods have to be employed in other cases, which was done in this work by using the COMSOL finite-element modeling (FEM) commercial software to compute the vibrational eigenmodes of coupled nanostructures. The case of coupled nanorods was considered first (paragraph b), as the low-frequency vibrational modes of cylindrical nanostructures with a length much larger than their diameter display an approximately longitudinal displacement field and are thus expected to present the same trends as in the slab case. The experimentally relevant case of a dimer of gold nanospheres coupled by a PVP spacer will be considered afterwards (paragraph c).

**b) Longitudinal vibrations of coupled nanorods**

The considered coupled nanorod system consists of two identical cylinders with a radius $R$ much smaller than their length $h$ ($R/h$=0.1 was used), connected by a cylindrical spacer of same radius $R$ and length $d$ (Fig. 2a left). The symmetries of the considered system were exploited by performing FEM simulations on a 2D system consisting of two rectangles of common width $R$ and lengths $h$ and $d/2$, respectively modeling one of the nanostructures and half of the spacer (Fig. 2a right). Indeed, the axial symmetry (about the $r$=0 axis) of the system allows one to use the axisymmetric version of the COMSOL Structural Mechanics module, greatly reducing the needed computation times as the problem to be solved becomes a 2D one instead of a 3D one. It should however be noted that this approach only enables to compute the subset of vibrational modes with a displacement field symmetric about the dimer axis (for spheres, such modes are indexed by a $m$ = 0 azimuthal number in Lamb's classification), 3D calculations being still required for other modes (as done e.g. in ref.[36]). Additionally, as the considered 2D system presents a reflection symmetry about the $z$=0 line (O$r$ axis in Fig. 2a right), its vibrational modes are necessarily either symmetric or antisymmetric about this axis (similarly to the slab model case). These two types of modes were thus successively computed by considering only half of the dimer in the FEM computations and imposing symmetry (antisymmetry) BCs on the $z$=0 axis, which corresponds to imposing a displacement direction on the $z$=0 axis parallel (orthogonal) to it.

The vibrational frequencies obtained for Au-PVP three-cylinders systems and for isolated gold nanostructures and PVP spacers are shown in Fig. 2b (information being presented in the same way as in Fig. 1b describing the slab case). Fig. 2b shows the first vibrational reduced eigenfrequencies (using the same reduced frequency $2fh/c_L$ as in the slab case, with $h$ representing the rod length here) computed for the symmetric (red lines) and antisymmetric modes (black lines) of PVP-coupled gold nanorod dimers. The reduced eigenfrequencies of isolated gold rods of length $h$, calculated using stress-free BCs all over the rod surface are shown by horizontal dashed blue lines. The reduced vibrational frequencies of isolated PVP rods of variable length $d$, calculated using displacement-free

BCs at the top and bottom nanorod ends ($z=\pm d/2$) and stress-free BCs on their lateral surface ($r=R$), are shown by dashed magenta (symmetric modes) and gray (antisymmetric modes) decaying curves.

The longitudinal vibrational frequencies of a free nanorod with large aspect ratio are approximately given by $f = \frac{n}{2h}\sqrt{\frac{Y}{\rho}}$, with odd and even $n$ corresponding to symmetric and antisymmetric modes, respectively.[44,45] This corresponds to reduced frequencies of $2fh/c_L = \frac{n}{c_L}\sqrt{\frac{Y}{\rho}} \approx$ 0.56 $n$ in the case of pure gold nanorods. Note that these reduced frequencies largely differ from the integer slab ones ; this is because the longitudinal deformation of rods with stress-free BCs at their lateral surface generates transverse strain because of Poisson effect, conversely to the slab case which corresponds to a purely uniaxial strain.[46] The lowest FEM-computed reduced frequencies of gold nanorods (with $R/h$=0.1) shown in Fig. 2b are in good agreement with this approximate formula. However, its validity decreases for increasing $n$, as diameter becomes increasingly comparable to longitudinal wavelength $h/n$, which explains the decreasing spacing between modes as frequency increases in Fig. 2b.

Fig. 2b present striking similarities with Fig. 1b, and all the qualitative trends observed in Fig. 1b and discussed in the previous paragraph remain observable in Fig. 2b. In particular, the vibrational frequencies computed for Au-PVP rods also form non-crossing branches which present a step-like evolution and are generally close to those obtained for isolated Au rods (with stress-free BCs on all surfaces) and PVP ones (with displacement-free BCs on the top and bottom surfaces and stress-free BCs at the $r=R$ one), deviating from them only near points where the frequencies of a nanostructure mode and that of a spacer mode cross each other, and in the quasi-contact regime. In the contact limit, the reduced frequencies of a free gold cylinder with length 2$h$, approximately equal to 0.28 $n$ for small $n$, are retrieved.

Fig. 2c presents the displacement fields computed (for three $d/h$ values) for dimer modes with reduced frequencies near $2fh/c_L \approx 0.56$, which corresponds to the frequency of the fundamental extensional mode of a free gold rod (whose displacement field is shown on the right of Fig. 2c), detectable in the context of optical time-resolved experiments.[39,45,47,48] Away from the quasi-contact regime and from crossing points (e.g. for $d/h$=0.5, green box in Fig. 2c), the extensional mode give rise to a symmetric dimer mode and an antisymmetric one with close reduced frequencies (≈0.56). For both of these dimer modes, displacement in the gold component is close to the displacement field of the fundamental mode of an isolated nanorod, and the two modes mostly differ (apart from their symmetrical/antisymmetrical character) by their associated displacement field in the PVP spacer. The grey box illustrates displacement fields obtained at the vicinity of a crossing point ($d/h$=0.92), corresponding to one of the $d/h$ values for which a mode of the PVP spacer (in this case an antisymmetric one with displacement-free BCs at both ends) presents a 0.56 reduced frequency identical to that of the fundamental extensional mode of a single gold cylinder. Avoided crossing of the antisymmetric PVP mode with the Au extensional mode leads in this case to two antisymmetric dimer modes with reduced frequencies of 0.50 and 0.61, respectively 10% smaller and higher than that of the gold nanorod extensional mode. A displacement amplitude larger in the PVP component than in the gold one and an assymetric displacement field in the latter one are observed in both cases. Modes obtained for $d/h$=0.02 are shown to illustrate the quasi-contact regime (blue box in Fig. 2c). While the antisymmetric dimer mode has a frequency and displacement field very close to those of the isolated nanostructure, the symmetric mode displays an increased frequency and a strong assymetry, displacement being much larger at the free end than at that in contact with the PVP spacer.

**c) Vibrations of coupled spheres**

The vibrations of a dimer of mechanically coupled spheres have already been the object of a few numerical analyses. Saviot and Murray considered both dimers made by overlapping spheres and dumbbell ones formed by spheres connected by a thin cylinder of much smaller radius.[22] In previous models of our inelastic scattering experiments on gold nanoparticles surrounded by PVP, whose morphology appears variable in electron microscopy images, we also considered the case of an infinite PVP matrix, which induces vibrational damping.[25,36]

We focus here on the complementary case of a cylindrical spacer with a radius close to that of the nanospheres. The dumbbell geometry considered for the FEM simulations is presented in Fig. 3a. It consists of two identical gold spheres with a diameter $D$, connected by a PVP spacer with a morphology more complex than in the slab and rod cases, but characterized by two parameters only, its diameter (a 0.95 $D$ value was used in the simulations, meshing being facilitated by the use of slightly different diameters for the nanosphere and the spacer) and length $d$, which was defined here as the smallest distance separating the two spheres. As in the case of rod systems, only modes involving a displacement field symmetric about the dimer axis were computed (corresponding to a $m = 0$ azimuthal number in Lamb mode classification[49]).

The lowest vibrational frequencies $2fD/c_L$ computed for the symmetric (red lines) and antisymmetric modes (black lines) of PVP-coupled gold nanosphere dimers are presented in Fig. 3b, an extended part of the eigenfrequency branch pattern being also shown in Fig. S10 of the Supporting Information. The reduced vibrational frequencies of free gold spheres of diameter $D$ (corresponding to Lamb modes with angular momentum and radial numbers values $l$ and $n$ indicated on the right of the plot) are also shown by horizontal dashed blue lines. Those of the PVP spacers, calculated using displacement-free BCs at both hemispherical spacer ends and stress-free BCs over the lateral cylindrical surface, are shown by horizontal dashed magenta (symmetric modes) and gray (antisymmetric modes) lines.

Similarly to the two previous systems considered, the vibrational frequencies computed for Au-PVP dumbbells are generally close to those obtained for the individual Au and PVP components of the full system, deviating from them only near points where the eigenfrequencies of the individual components approach and cross each other, and in the quasi-contact regime (Fig. 3b). In the latter regime, the frequencies of symmetric modes all increase with decreasing $d/D$, but in a strongly mode-dependent way. For instance, the frequency of the symmetric mode deriving from the translation Lamb mode ($l = 1$, $n = 0$ mode with $2fD/c_L$=0) is strongly affected by $d/D$ decrease, while that of the symmetric mode deriving from the ($l = 1$, $n = 1$) Lamb mode ($2fD/c_L$=0.79) is almost unaffected by $d/D$. The symmetric mode deriving from the fundamental quadrupolar Lamb mode ($l = 2$, $n = 0$, $2fD/c_L$=0.55) presents an intermediate behavior, its frequency increasing of up to 10 % in the quasi-contact regime.

Fig. 3c presents displacement fields computed for dimer modes with reduced frequencies close to that of the fundamental quadrupolar mode of isolated gold spheres, which dominates their inelastic light scattering spectra.[2,33] Coupling effects similar to those illustrated by Fig. 2c in the coupled rod case are observed. Away from the quasi-contact regime and from crossing points (e.g. for $d/D$=0.3, green box in Fig. 2c), the symmetric and antisymmetric dimer modes deriving from the quadrupolar sphere mode (whose displacement field is shown on the right of Fig. 3c) display similar eigenfrequencies and displacement fields in the gold component. Conversely, these features are modified at the vicinity of crossing points (e.g. for $d/D$=0.64, grey box in Fig. 3c). As in the rod dimer case, avoided crossing of an Au mode and an antisymmetric PVP one leads to two antisymmetric dimer modes with different (0.51 and 0.59) reduced frequencies, both presenting a displacement amplitude larger in the PVP component than in the gold one, as well as an assymetric displacement field in gold. In the quasi-contact regime (blue box in Fig. 2c, corresponding to $d/D$=0.02), the quadrupolar sphere mode produces both an antisymmetric dimer mode with a similar frequency and displacement field and a symmetric one with a higher frequency and a modified, strongly assymmetric field.

The quasi-contact limit is more complex for sphere dimers than for coupled slabs and cylinders, and cannot be related to the vibrational modes of a pure gold nanostructure, because $d$=0 corresponds to a situation where PVP is still present. Nevertheless, the frequency increase of symmetric modes in the quasi-contact regime limit can still be qualitatively interpreted in terms of a modification of the BCs at the surface of the gold spheres induced by the presence of the spacer. This is demonstrated by Fig. 4, which presents the vibrational frequencies of a gold sphere calculated using hybrid BCs on the sphere surface, with displacement-free conditions on a cone of half-angle $\theta$ and stress-free BCs elsewhere (Fig. 4a). An increase of $\theta$ is seen to induce a strongly mode-dependent increase of the vibrational frequencies of the sphere (Fig. 4b), which is the same effect as a decrease of $d/D$ for the symmetric modes of the Au-PVP system (Fig. 3). For instance, $\theta$ increase is shown to induce strong, weak and intermediate frequency shifts for the ($l$ = 1, $n$ = 0), ($l$ = 1, $n$ = 1) and ($l$ = 2, $n$ = 0) Lamb modes, respectively, which correspond to the trend observed when decreasing $d/D$ values in the quasi-contact regime for Au-PVP dimers (Fig. 3). Fig. 4c shows how the displacement fields of Lamb modes (corresponding to $\theta$=0°) relevant for inelastic light scattering and time-resolved experiments on dimers[21,25,36] (fundamental $l$ = 0, 1 and 2 modes, corresponding to breathing, translation and quadrupolar Lamb modes) evolve when $\theta$=16° (this value having been chosen as it leads to vibrational eigenfrequencies close to those obtained for the Au-PVP system in the contact case). For all modes, displacement fields become much less symmetric as in the $\theta$=0° case, as expected from the symmetry loss induced by the hybrid BCs used in this case.

The main difference between the calculations reported here (involving nanospheres coupled by a finite spacer) and those that we previously reported for quasi-translation and quadrupolar modes, in which an inifinite PVP matrix was considered,[25,36] is the existence of damping in the latter case, associated with the emission of acoustic waves in the matrix. Apart from this difference, the vibrational frequencies computed for nanosphere dimers in the framework of the finite spacer and infinite matrix models do not differ much. For instance, reduced frequencies of 0.20 and 0.62 were computed in the contact case for the axisymmetric symmetric quasi-translation and quadrupolar dimer modes, respectively,[36] to be compared with the 0.17 and 0.61 obtained here. The small differences can be ascribed to a slightly stronger coupling for an infinite matrix than for the finite spacer configuration. In both cases, the computed reduced frequencies are a bit smaller than the experimentally measured ones (0.27 and 0.71),[36] which probably results from the simplifications made for carrying out the FEM simulations (electron microscopy showing in particular that the nanoparticles used were not perfectly spherical).

**CONCLUSIONS**

We have computed the low-frequency acoustic eigenmodes of dimers of identical nanostructures with different morphologies (slabs, rods and spheres) mechanically coupled by a spacer, and compared them to the eigenmodes of their individual components. In the investigated situations, corresponding to a homodimer of gold nanostructures coupled by a PVP spacer (leading to modes of the Au-PVP composite system either symmetric or antisymmetric), coupling effects are moderate due to the large acoustic impedance mismatch between gold and PVP, and the vibrational frequencies of the coupled systems are in most cases close of those computed for their individual components using relevant BCs. Nevertheless, mechanical coupling affects the eigenmode features (frequency and displacement field) in two situations when the spacer thickness is varied: in the near-contact regime, where the symmetric modes of the composite Au-PVP system present eigenfrequencies largely modified as compared to those of the individual gold components, which can be qualitatively interpreted as a partial change of BCs on the surface of the metal nanostructures, and when two modes of the individual components present identical frequencies, in which case avoided crossings occur in the eigenfrequency branch pattern. Whereas the frequency shifts occurring in the near-contact regime between PVP-coupled gold nanostructures have already been experimentally observed, demonstrating avoided crossing effects is more challenging. It would require an excellent control and precise knowledge of the morphology of the investigated nanosystems, and would be facilitated if the acoustic impedance mismatch between the nanostructures and their spacer was reduced as compared to the Au-PVP case considered here. The sensitivity of the symmetric eigenmodes of metal nanostructure dimers to the thickness and mechanical properties of the spacer makes vibrational measurements a potentially useful tool to characterize nanometric spacers. Future work may involve more systematic experimental investigations of vibrational frequencies as a function of separating distance, as well as the design of more sophisticated analytical models to address the vibrations of dimers or more complex oligomers of nanospheres, which could be done by transferring the concepts underlying generalized Mie theory, used for the calculation of the electromagnetic response of coupled spheres, to the acoustics field.

**Corresponding author**

**\***E-mail: aurelien.crut@univ-lyon1.fr

**ORCID**

Jean Lermé :  0000-0002-4094-5100

Jérémie Margueritat : 0000-0003-2075-1875

Aurélien Crut: 0000-0003-2185-709X

**Notes**

The authors declare no competing financial interest.

**ASSOCIATED CONTENT**

**Supporting information**

> Slab model : single slab and homodimer cases, results, hybridization analysis, dependence of coupled vibrational modes on the impedance acoustic ratio *p*.

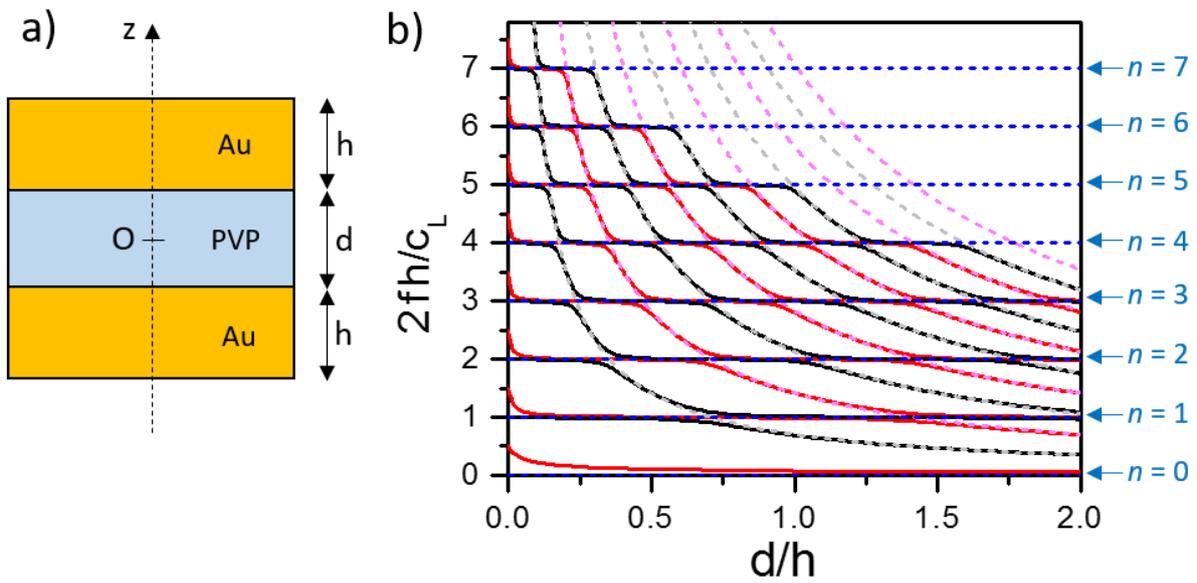

**Figure 1. Vibrations of coupled slabs.** a) Geometry considered in the calculations. It consists of two Au slabs of thickness h (with infinite lateral extension) separated by a PVP slab of thickness d. b) Vibrational frequencies computed using the analytical model described in the main text and in the Supporting Information for the symmetric (red lines) and antisymmetric modes (black lines) of the three-slabs system. The frequencies of isolated slabs (computed using stress-free and displacement-free BCs for Au and PVP slabs, respectively) are shown as dashed blue horizontal lines (gold slab modes, indexed by an integer $n$ shown on the right) and dashed magenta/gray lines (symmetric/antisymmetric modes of the PVP spacer). Note that a limited number of branches are plotted for the sake of clarity.

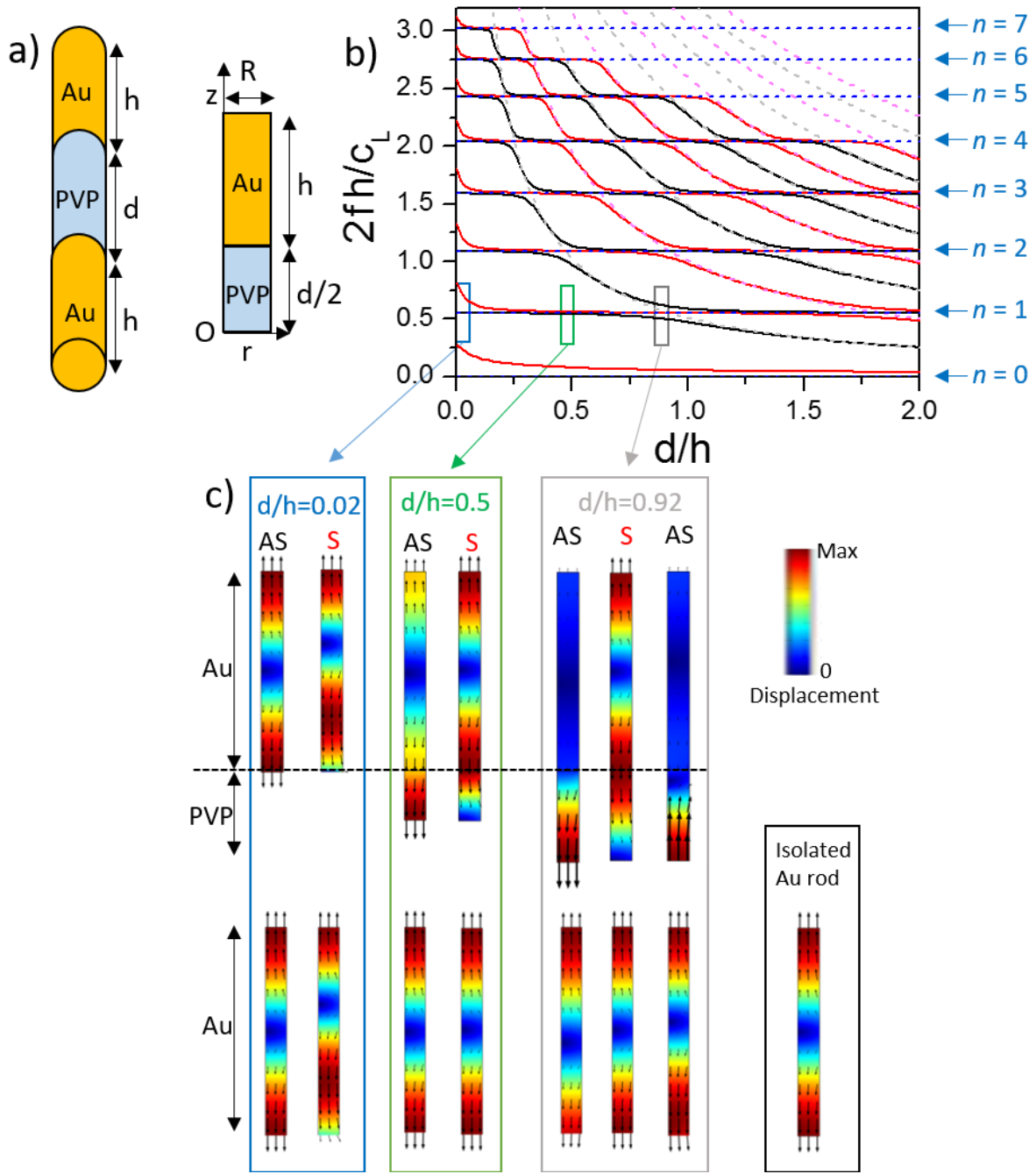

**Figure 2. Vibrations of coupled rods.** a) 3D geometry of the considered coupled rod system (left) and 2D one used for FEM computations of its axisymmetric vibration modes (right). b) Vibrational eigenfrequencies computed using FEM for the symmetric (red lines) and antisymmetric modes (black lines) of gold dimers. The eigenfrequencies of isolated rods are shown as dashed blue horizontal lines (gold rod modes, computed with stress-free BCs) and dashed magenta/gray lines (symmetric/antisymmetric modes of the PVP spacer, computed with displacement-free BCs at both plane ends and stress-free BCs over the lateral cylindrical surface). c) Associated displacement of the modes contained in the boxes shown in b, corresponding to reduced frequencies $2fh/c_L \approx 0.55$ and to $d/h=0.02$, 0.5 and 0.92. In all cases, a 0-Max scale is used for plotting the displacement fields. The displacement in one half of the dimer is shown on top (geometry of Fig. 2a right), while the displacement in the gold component is shown on the bottom, the different scale used facilitating the comparison between the different situations. The displacement field of a gold cylinder with stress-free BCs is also shown on the right.

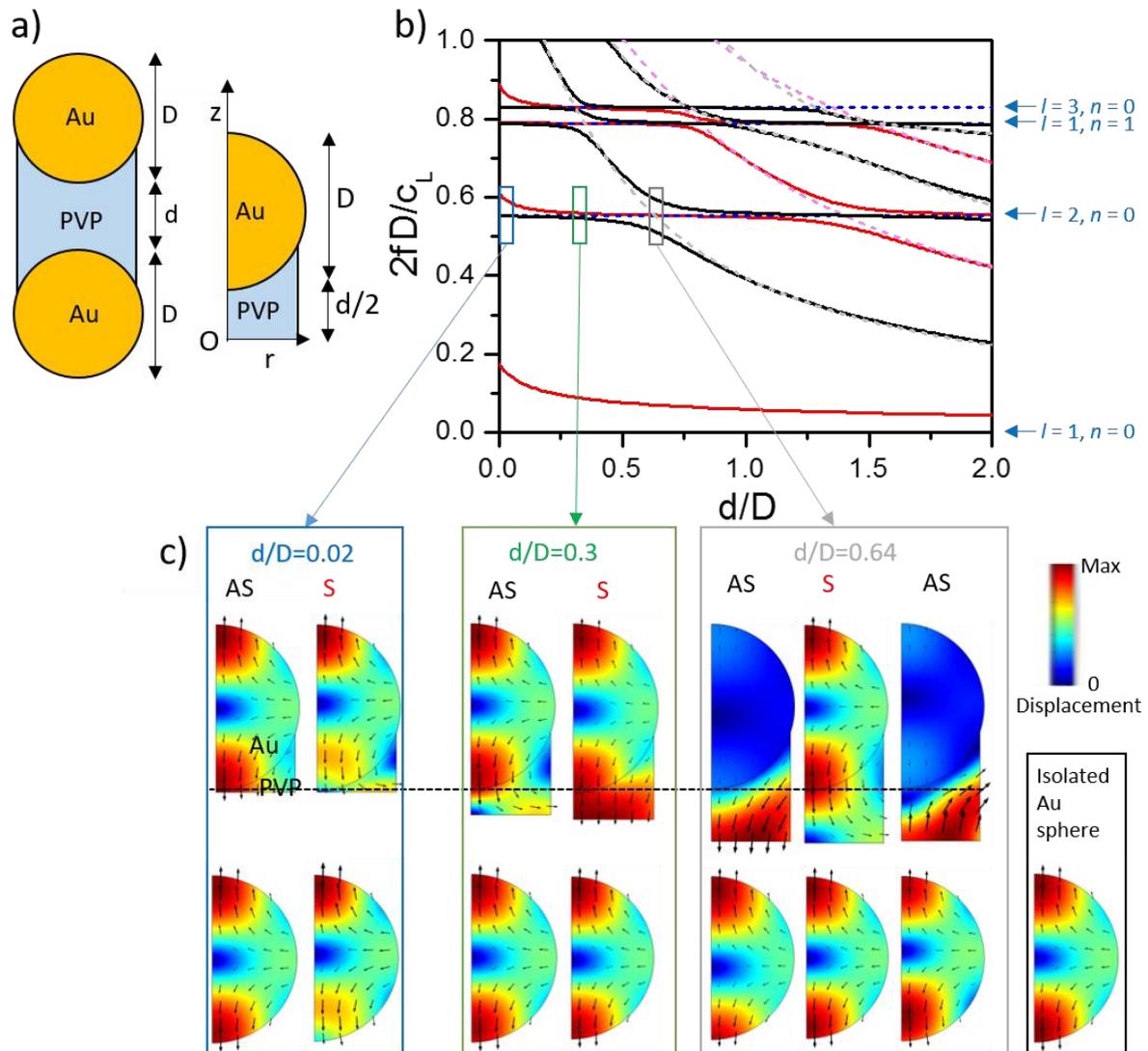

**Figure 3. Vibrations of coupled spheres.** a) 3D geometry of the considered coupled sphere system (left) and 2D one used for FEM computations of its axisymmetric vibration modes (right). b) Vibrational eigenfrequencies computed using FEM for the symmetric (red lines) and antisymmetric modes (black lines) of gold dimers. The eigenfrequencies of isolated components are shown as dashed blue horizontal lines (gold sphere Lamb modes, computed with stress-free BCs on the sphere surface) and dashed magenta/gray lines (symmetric/antisymmetric modes of the PVP spacer, computed with displacement-free BCs at the spacer hemispherical ends, and stress-free BCs on its lateral cylindrical surface). c) Associated displacement of the modes contained in the boxes shown in b, corresponding to reduced frequencies $2fD/c_L \approx 0.55$ and to d/D=0.02, 0.3 and 0.64. The displacement field of the fundamental quadrupolar mode of a gold sphere with stress-free BCs is also shown on the right.

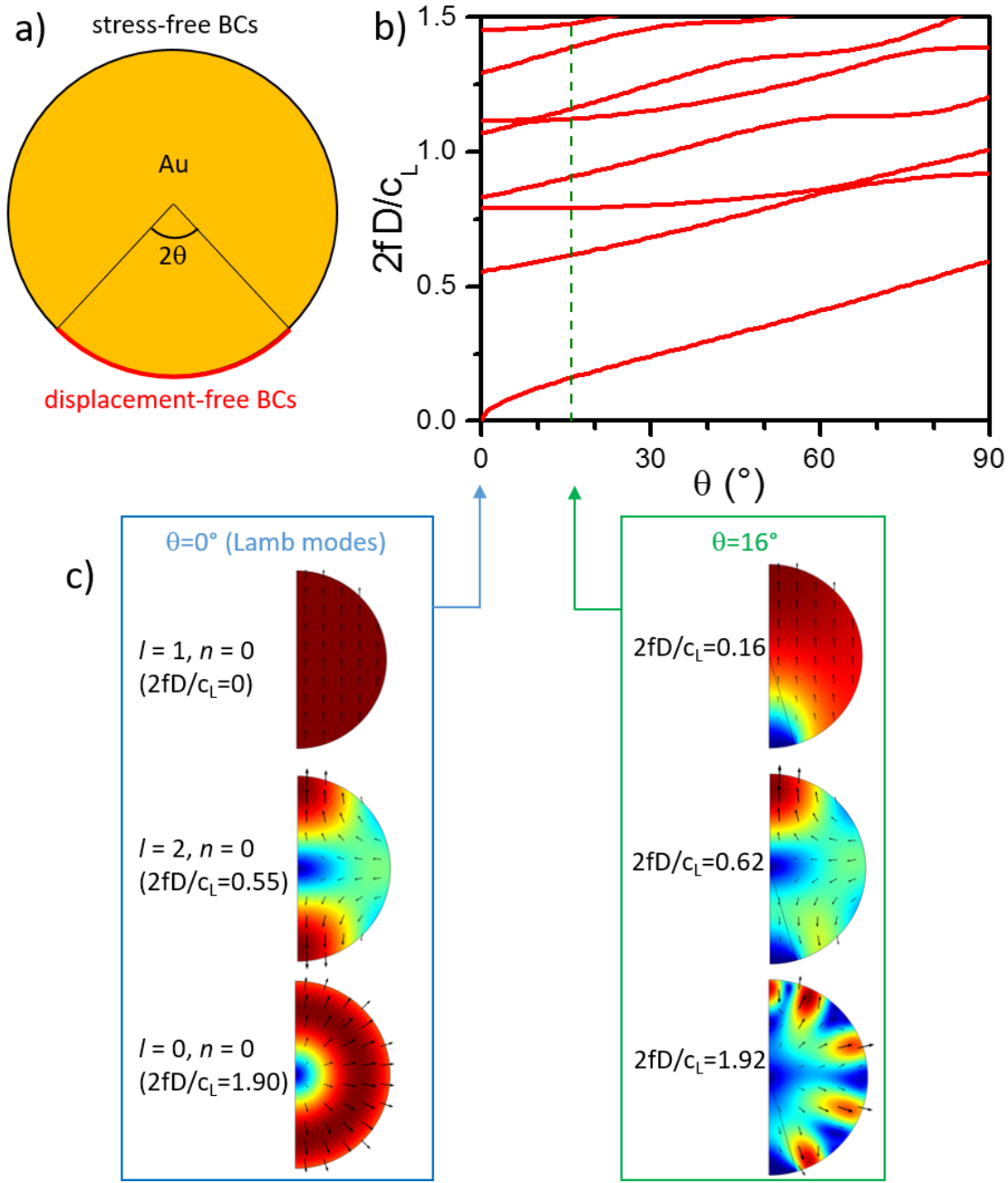

**Figure 4. Vibrations of a single gold sphere with hybrid BCs.** a) Considered situation : gold sphere with displacement-free BCs on a cone of half-angle θ and stress-free BCs elsewhere. b) FEM-computed reduced frequencies. c) Displacement fields computed for θ=0° (Lamb modes) and θ=16°.

**TOC**

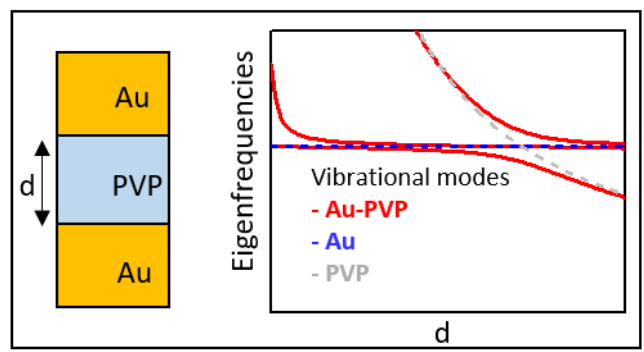